# Self-powered smart contact lenses: a multidisciplinary approach to micro-scale energy and 900 MHz – 1.1 GHz bandwidth microfabricated loop antennas communication systems

Patrice Salzenstein [1*], Blandine Guichardaz [1], Aya M. Bessou [1], Ekaterina Pavlyuchenko [2], Martine Comte [3] and Maxim V. Pogumirsky [4,5]

[1] Centre National de la Recherche Scientifique (CNRS), Université Marie et Louis Pasteur (UMLP), FEMTO-ST Institute, 15B Avenue des Montboucons, F-25030 Besançon Cedex, France ; patrice.salzenstein@cnrs.fr ; blandine.guichardaz@femto-st.fr ; ayamaroua.bessou@femto-st.fr

[2] Centre National de la Recherche Scientifique (CNRS), F-91190 Gif-sur-Yvette, France ; ekaterina.pavlyuchenko@cnrs.fr

[3] Consultant, Canton of Neuchâtel, CH-2300 La Chaux-de-Fonds, Switzerland ; martine_comte2003@hotmail.com

[4] FAREXPORT Ltd., 190005, Saint Petersburg, Russia; max.pogumirsky@gmail.com

[5] National Research University of Information Technologies, Mechanics and Optics (ITMO), Saint Petersburg, Russia

* Correspondence: patrice.salzenstein@cnrs.fr

**Abstract:** Smart contact lenses are at the forefront of integrating microelectronics, biomedical engineering, and optics into wearable technologies. This work addresses a key obstacle in their development: achieving autonomous power without compromising safety or miniaturization. We examine energy harvesting strategies using intrinsic ocular sources—particularly tear salinity and eyelid motion—to enable sustainable operation without external batteries. The study emphasizes compact loop antennas operating between 900 MHz and 1.1 GHz as critical for wireless data transmission and power management. Material choices, signal integrity, and biocompatibility are also discussed. By presenting recent advances in 3D-printed optics, antenna integration, and energy systems, we propose a conceptual framework for the next generation of smart lenses, enabling real-time health monitoring and vision enhancement through self-powered, compact devices.

**Keywords:** Smart contact lenses; tear salinity; eye-blink energy; micro-batteries; biocompatibility; loop antenna

## 1. Introduction

Since the invention of contact lenses by Adolf Gaston Eugen Fick in 1888 [1], their primary purpose has been vision correction.

Advantages of Contact Lenses are well known. Among the advantages are: (1) Unobstructed peripheral vision, unlike glasses. (2) No glare or visual distortion. (3) Freedom during sports or leisure activities. (4) Aesthetically, they do not alter appearance, unlike glasses. These advantages explain why many people prefer contact lenses for various daily activities.

However, there are still Potential Problems and Disadvantages, with some limitations (5) Risk of eye infection due to poor hygiene. (6) Dryness and irritation at the end of the day if the tear film is insufficient. (7) Requires a strict cleaning and replacement routine. (8) Initial adaptation difficulties for some sensitive users.

New possibilities could really bring additional advantages and be a kind of revolution in the way of considering contact lenses.

In the last decade, however, the landscape has evolved with the introduction of smart contact lenses [2–4], designed to offer functionalities ranging from augmented reality displays to real-time biosensing. More precisely, Robert *et al* [5] show that a scleral contact lens with integrated near-infrared lasers – Contact Lens Pointer (CLP) significantly enhances eye-tracking accuracy and reliability in challenging lighting conditions, especially outdoors, compared to a commercial eye tracker. Robert *et al* [6] introduces an infrared laser pointer integrated into a scleral contact lens, powered wirelessly and equipped with a diffractive optical element for beam collimation or image projection, detailing its design, assembly challenges, and potential applications in visual assistance and augmented reality.

There are papers in recent literature that show the interest and the possibility of making smart lenses. Smart contact lenses (SCLs) are emerging as innovative tools for non-invasive monitoring of blood sugar, intraocular pressure, and other health conditions, potentially revolutionizing the management of diseases like glaucoma [7]. Contact lenses are emerging as a powerful non-invasive platform for continuous, multi-biomarker sensing in tear fluid, though challenges like stability, manufacturing, and regulation must be overcome for clinical use [8]. Some recent advances in wearable medical contact lenses, focusing on their applications in intraocular pressure monitoring, glucose detection, ocular drug delivery, and color blindness treatment, with special emphasis on the integration of electro spun fibers for controlled drug release, current challenges, and future opportunities for real-world implementation [9]. Kim *et al* discusses the latest advancements in smart contact lens systems for ocular drug delivery and therapy, highlighting their non-invasive benefits, drug and light delivery capabilities, various design approaches, and future prospects [10].

A fundamental challenge in realizing these innovations lies in providing reliable, autonomous power to microscale electronic systems embedded within the lens. Two unconventional yet promising approaches have been proposed: (i) harvesting mechanical energy from involuntary eyelid movements (blinking), and (ii) exploiting the ionic content of tears as an electrochemical power source [11]. In Singapore, Yun *et al* have developed bio chargeable, tear-based batteries embedded in smart contact lenses that utilize enzymatic glucose reactions for safe and efficient power generation during storage, enabling both biological and conventional charging [12]. The same group, with Li *et al*, has developed a power-free smart contact lens that can noninvasively monitor glucose levels through color changes in electrochromic electrodes, eliminating the need for external electronics and enabling cost-effective, daily health monitoring [13].

Lens design depends on several factors related to the required correction and desired comfort:
- Spherical: Corrects myopia and hyperopia with a uniform curvature.
- Toric: Designed specifically to correct astigmatism, they have different powers in different meridians of the lens.
- Multifocal: Used for presbyopia, allowing near and distance vision thanks to zones of varying power.

Building on our prior work in optics and lens manufacturing [14,15], this article aims to present the current state of the art and discuss the potential energy sources for powering these next-generation devices.

The manuscript is structured into several key sections. It begins by outlining the functional principles of smart contact lenses, followed by an exploration of autonomous powering strategies using eye-intrinsic and alternative energy sources. Subsequent sections address the technical expertise required and design constraints involved in lens development, as well as the operating frequency range relevant to integrated antenna systems. The paper concludes with a synthesis of findings and perspectives on future development.

## 2. Smart Contact Lenses: functional principles

Smart contact lenses integrate sensors, communication interfaces, and power management circuits into a transparent, soft substrate worn directly on the cornea. Potential use cases include:

*2.1. Non-invasive health monitoring (e.g., glucose or intraocular pressure)*

Non-invasive health monitoring refers to the use of technology to track health indicators like blood glucose or intraocular pressure without breaking the skin or entering the body. Instead of using needles or surgical tools, these methods rely on sensors, light-based technologies, or wearable devices to gather data through the skin or eyes. For example, some smartwatches can estimate glucose levels using optical sensors, and smart contact lenses are being developed to monitor eye pressure in glaucoma patients. It's a growing field aiming to make health tracking easier, safer, and more comfortable.

*2.2. Augmented vision (e.g., night vision, real-time text translation)*

Augmented vision refers to technologies that enhance human visual perception by overlaying or enhancing visual information. Examples include night vision, which amplifies low light to help see in darkness, and real-time text translation, which uses augmented reality to display translated text over signs or menus through devices like smart glasses or phone cameras. These tools expand what the human eye can perceive or understand in real time.

*2.3. Visual aids for the visually impaired*

Visual aids for the visually impaired are tools and technologies designed to help people with partial or full vision loss better perceive their surroundings, access information, and perform daily tasks. Their main goal is to promote independence and improve quality of life through enhanced accessibility. Additionally, SLCs can take over the convergence power of the lens, in order to correct defects that can affect the ciliary muscle. The ciliary muscle is an ocular smooth muscle belonging to the ciliary body whose role is to modulate the convergent power of the lens and to allow accommodation in the field of vision.

Thiele *et al* presents the first demonstration of 3D-printed hybrid refractive/diffractive achromatic and apochromatic micro lenses using multi-material direct laser writing, significantly reducing chromatic aberrations in the visible range by combining diffractive surfaces and photoresists with different dispersions [16]. The same group 3D-printed a compact, multi-lens foveated camera system directly onto a CMOS chip, mimicking eagle vision by combining lenses with varying focal lengths to enhance central image resolution [17].

Interesting functionalities can be achieved by working with materials such as Polydimethylsiloxane (PDMS) for 3D-printed lenses. PDMS an elastomer with excellent optical, electrical and mechanical properties, which makes it well suited for several engineering applications. Due to its biocompatibility, PDMS is widely used for biomedical purposes.

Recent studies have looked into adding smart features to contact lenses, focusing on things like adaptive vision correction and real-time changes in optics [18–20]. Early designs mostly used soft, flexible materials that work well with the eye and offer decent optical performance [21]. More recently, advanced 3D printing methods, including two-photon polymerization, have given researchers new ways to create tiny, precise structures and complex lens shapes [22]. This shift in technology is changing how optical parts are made for biomedical uses, especially in eye care, where 3D-printed components support more sophisticated and responsive visual systems [23]. Building on these developments, our research aims to develop new adaptive optical parts for contact lenses using the latest microfabrication tools.

*2.4. Common considerations*

The design must balance miniaturization, transparency, oxygen permeability, and mechanical comfort. The integration of components integrated into the lens is shown schematically in Figure 1.

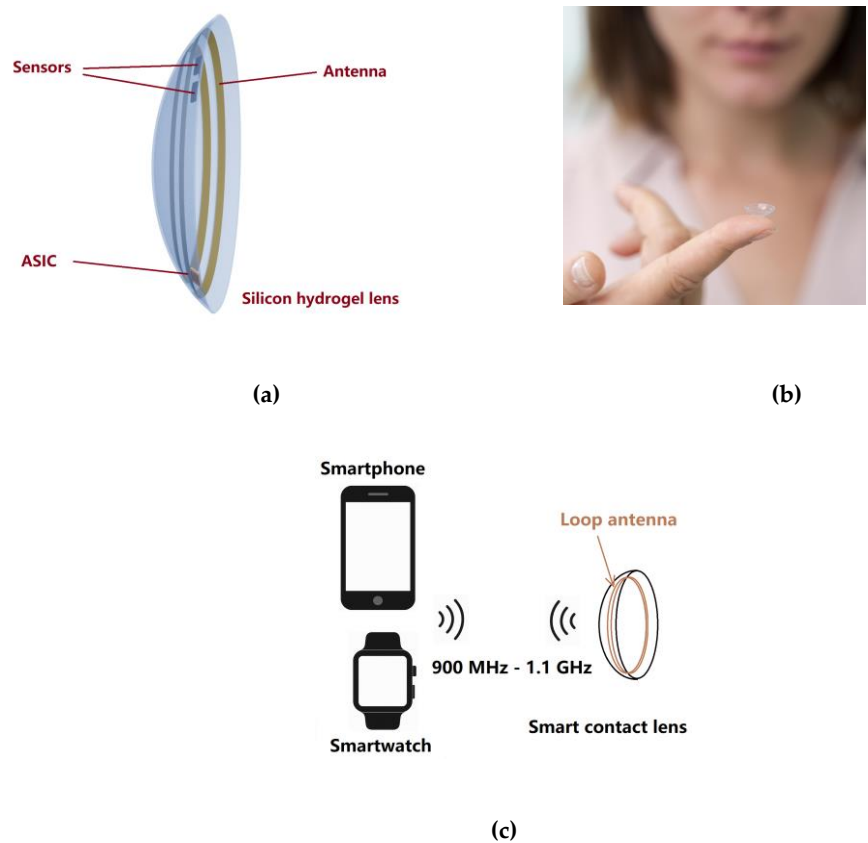

**Figure 1. (a)** This Figure illustrates the general principle of making a contact lens, which includes an antenna, an application-specific integrated circuit (ASIC), and sensors to communicate with a smartwatch, a mobile phone, or a computer. Technologically, the addition of components in a lens allows, if necessary, to position a progressive contact lens by correctly orienting it. These technologies are feasible. **(b)** This Photography [24] illustrates the scale and delicacy of a standard contact lens, highlighting the complexity of embedding electronic systems within such a confined space. **(c)** Illustration of the communication between a smart contact lens and a smartwatch and/or a smartphone. Indicated frequency bandwidth corresponds to what is discussed in Part 5 of this paper.

Advanced materials such as flexible biocompatible polymers and nanoscale thin films are crucial to achieving these goals. Recent reviews have discussed material advances and sensor integration for ophthalmic applications. Smart contact lenses are emerging wearable devices that enable non-invasive disease monitoring and therapy by detecting various ocular biomarkers and delivering drugs, although their development is still in early stages [25]. Smart contact lenses are being actively developed for non-invasive health monitoring by leveraging the eye's ability to reveal both physical and chemical biomarkers, with recent advances focusing on biosensing, data transmission, drug delivery, and energy storage technologies [26]. Tear exchange between the ocular surface and contact lenses is limited, especially with soft lenses, leading to debris build-up and potential complications, and despite innovations in lens design and measurement methods like fluoro-photometry, understanding of tear hydrodynamics and exchange remains insufficient [27].

## 3. Autonomous powering via eye-intrinsic or other sources

This development of smart contact lenses has accelerated, driven by advancements in bio-compatible electronics and the growing need for autonomous wearable devices. A key challenge remains: how to provide sustainable, safe, and efficient power to these miniature systems.

This section explores emerging strategies for energy harvesting directly from the human body or its immediate environment. Section 3.1 investigates the use of tear salinity as a natural electrolyte for powering biofuel cells. Section 3.2 focuses on biomechanical energy generated from blinking, leveraging nanogenerators to convert motion into electricity. Lastly, Section 3.3 considers other potential sources, including kinetic energy from head and body movements, offering new avenues for self-sustaining smart lenses, and Section 3.4. discuss aspects of integration.

*3.1. Energy from tear salinity*

Tears contain a mixture of electrolytes—mainly sodium and potassium ions—that can be harnessed for low-power biofuel cells. Recent developments have demonstrated tear-based micro-batteries that operate continuously by leveraging this natural saline environment as discussed in the introduction. In particular, glucose-coated ultra-thin batteries can undergo redox reactions catalyzed by tear fluid, extending the lens's operational time.

It has to be underlined that in Singapore, the group of Seok Woo Lee from Nanyang Technological University (NTU), demonstrated a prototype contact lens with such a battery, capable of lasting 13 hours using only tear fluid as an energy source [12,13,28], where it is shown how recent advancements in smart contact lenses (SCLs) have led to innovative features like overlaying information and monitoring blood glucose levels, but these require a reliable power source. Lithium-ion batteries are currently used, but due to safety concerns, polymerized hydrogels like Copper Hexacyanoferrate (CuHCFe) and Prussian Blue (PB) – or *Preußischblau* in German (used in medicine as an antidote for certain kinds of heavy metal poisoning and for electrochemical energy storage) [29], are being explored as safer alternatives, with promising results showing a storage capacity of 0.132 mAh (the amount of current that a battery can supply for one hour before it is fully discharged), when tested with tear solutions. The lens remains fully biocompatible and safe for prolonged use.

*3.2. Energy from blinking*

An alternative energy harvesting method relies on biomechanical motion. Eyelid blinking, occurring tens of thousands of times daily, can actuate piezoelectric or triboelectric nanogenerators. Innovative hybrid systems combining a flexible silicon photovoltaic cell with a blink-activated Mg–$O_2$ collector have been proposed [30], enabling dual-mode energy capture (optical and mechanical): this paper discusses a hybrid energy generation system for powering smart ocular devices, combining a flexible silicon solar cell and an eye-blinking activated Mg–$O_2$ metal–air harvester. The system continuously generates electrical power, providing stable DC output without needing external accessories, with power management circuits boosting voltages for consistent energy supply to ocular devices.

Such energy harvesting solutions pave the way for fully autonomous, wire-free lenses that recharge "in the blink of an eye."

*3.3. Other potential sources*

It's possible to imagine other energy sources that could help smart lenses function. Natural head movements or the movements of an individual walking or running are possible potential energy sources that could be used to drive smart lenses. We know that the human body involves an expenditure of energy which can also be approached in frequency terms [31]. What is generally perceived as an energy cost could turn out to be a source of energy for microdevices. Utilization of elastic energy in human movement is definitely a possible source for micro batteries [32].

*3.4. Integration*

Figure 2 illustrates how the integration of a miniaturized energy source obviously requires specific design work adapted to a contact lens.

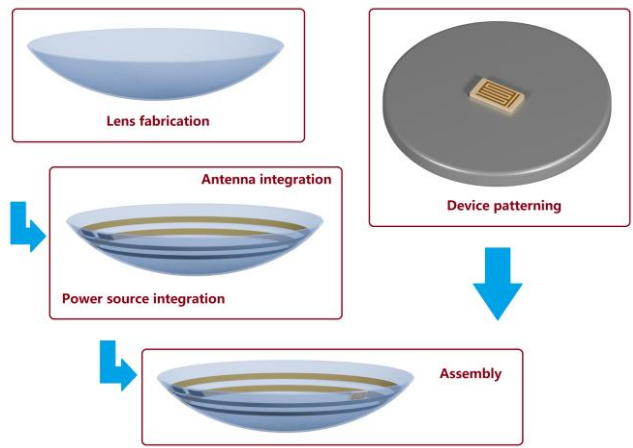

**Figure 2.** This Figure illustrates the conceptual roadmap for integrating multiple energy harvesting methods into 10 mm diameter smart lenses.

Integrating multiple energy harvesting methods into 10 mm smart contact lenses demands such a conceptual roadmap. This includes optimizing a miniaturized energy source through hybrid systems — Power source — while maintaining biocompatibility. A 9.5 mm double-loop antenna must be precisely deposited on the periphery of the lens. A low power ASIC manages power distribution and sensor data. The integration of two or more sensors requires careful spatial design to ensure seamless functionality, user comfort, and reliable wireless communication within the compact lens architecture. In addition, antennas and active or passive components must be protected by a thin layer ensuring the passage of oxygen and biocompatibility. For 3D-printed lenses, manufacturing processes become more complex. We remind that the weight asymmetry due for example to ASIC is a guarantee of self-alignment for lenses correcting astigmatism or presbyopia.

## 4. Technical expertise and design constraints

The development of smart lenses—devices capable of integrating various technologies for visual enhancement or information display—is a highly complex challenge. The feasibility of creating these lenses hinges on expertise in multiple technical domains. Here is a deeper look into the Following five points:

• <u>Fabrication of thin-film optical coatings related to our experience in lens manufacturing</u>: The fabrication of thin-film optical coatings [14,15] is crucial for controlling the behavior of light passing through or reflecting off the lens surface. These coatings can serve various purposes, such as enhancing the optical quality of the lens, providing anti-reflective properties, or enabling the lens to interact with light in specific ways (e.g., filtering, polarization control). Achieving precise and high-quality fabrication of these coatings is essential to ensure that smart lenses not only meet optical requirements but also maintain durability, especially in demanding environments like every day wear or medical applications.

• <u>Analysis of material interfaces for biocompatible coatings and signal transfer – Specific Absorption Rate (SAR)</u>: The interaction between materials used in the smart lens must be carefully considered, particularly in areas where the lens will come into direct contact with human tissues or fluids, such as the eye. Biocompatible coatings are necessary to prevent irritation, inflammation, or other adverse effects. These coatings should not only be chemically compatible with the tissues of the eye but also enable efficient signal transfer (such as through wireless communication or light signals) without degrading over time. Understanding and analyzing material interfaces [33–37] is critical because any imperfections in these interfaces could lead to issues like reduced optical performance, unreliable data transfer, or irritation. Research in biocompatible coatings and materials, therefore, plays a significant role in ensuring that smart lenses are safe, functional, and long-lasting. For individuals using SLCs, it's crucial to ensure that the magnetic loop antenna's emitted power remains within safe limits to prevent potential harm to the human body. Regarding Specific Absorption Rate (SAR) standards, SAR measures the rate at which the body absorbs energy from radio frequency electromagnetic fields, expressed in watts per kilogram (W/kg). Various international standards define exposure limits to ensure safety. For example, the European Union sets a SAR limit of 2.0 W/kg averaged over 10 grams of tissue [38,39]. It is legally and medically reasonable to follow the precautionary principle: for SAR, there is no standard that specifically concerns a device or a component that would be in contact with the eye, which is why it is prudent to refer to the two standards *IEC/IEEE 62209-1528:2020* [38] and *IEC 62209-3:2019* [39] that concern SAR of human exposure to radio frequency fields from hand-held and body-mounted wireless communication devices, in particular *IEC 62209-3:2019*. In the United States, the Federal Communications Commission mandates a SAR limit of 1.6 W/kg averaged over 1 gram of tissue for mobile devices. In China, the maximum permitted SAR value for mobile phones is 2.0 W/kg, measured over 10 grams of tissue, with the same standard used in Europe, based on the International Commission on Non-Ionizing Radiation Protection (ICNIRP) guidelines. In India, the SAR limit for mobile devices is 1.6 W/kg. More generally, the World Health Organization regularly updates the SARs by country, with its "Exposure limits for radio-frequency fields (public) – Data by country" [40].

• Frequency analysis and signal integrity diagnostics in systems: Smart lenses integrate systems that use light (optical signals) to perform tasks like image recognition, display projection, or health monitoring. The performance of these systems depends heavily on maintaining the integrity of the signals being transmitted and received. Any disruption in the signal (due to noise, interference, or attenuation) can significantly impair the functionality of the lens. Frequency analysis is vital to understanding how signals behave [41,42] in the components of the lens, especially in the context of wireless communication (e.g., 2400 – 2483.5 MHz Bluetooth, 2.4 GHz Wi-Fi for 450 Mbps or 600 Mbps, 5 GHz Wi-Fi for up to Gbps, 900 MHz 3G, 1800 4G, and 3.5 GHz 5G microwaves for upper bandwidth) and power transfer. Signal integrity diagnostics are necessary for detecting and troubleshooting issues in signal transmission, ensuring reliable operation of the smart lens over time. Ensuring that these components operate seamlessly is crucial for the lens to function as intended. In France, for example, the distribution of frequency bands is codified and distributed by the National Frequency Agency. The National Frequency Band Allocation Table is set by a decree of the Prime Minister [43]. The frequency band likely to be used for communications between SLCs and a smartphone or smartwatch is in the band around 1 GHz, as we see in Part 5. It is therefore important to control the risks associated with potential intermodulation effects. This problem has been the subject of studies, both regarding passive intermodulation distortion in antennas [44], and for the interaction with amplifier elements in electronics [45–47]. Just above 1 GHz, another widely used application is Global Navigation Satellite System (GNSS) [49–52], with bottom of the lower Band-L (1.151–1.214 GHz) for GPS L5 and Galileo E5 are located, with E5a and L5 coexisting in the same frequencies. The remaining L2 (GPS), G2 (GLONASS) and E6 (Galileo) signals are in the 1.216–1.350 GHz bandwidth. These frequencies are given in Table 1. These bands are allocated to Radio-location Services for ground radars by the Aeronautical Radio Navigation Service and Radio Navigation Satellite System on a primary basis, thence the signals in these bands are more vulnerable to interference transmission and reception signals up to 1.1 GHz.

**Table 1.** Operating frequency range of Global Navigation Satellite System below 1.3 GHz[1,2].

| Attributed band designation (Centered frequency in GHz) | Satellite Network[2] | Range (GHz) |
|---|---|---|
| E5a/b | Galileo | 1.164–1.215 |
| L5 (1.176) | GPS, NavIC, QZSS | 1.215–1.234 |
| G3 | GLONASS | 1.189–1.214 |
| B2 (1.207) | BeiDOU | |
| L2 and L2C (1.228) | GPS and QZSS | 1.215–1.234 |
| B3 (1.268) | BeiDou | |
| G2 | GLONASS | 1.237–1.254 |
| L6 (1.229) | QZSS | |
| E6 | Galileo | 1.260–1.300 |

[1] Those GNSS also operate at upper frequency range but these frequencies are not mentioned as they are not supposed to interact with the 900 MHz – 1.1 GHz bandwidth.
[2] Signal-in-space ranging error were in 2019 in the range of 1.6 cm for Galileo, 2.3 cm for GPS, 5.2 cm for GLONASS and 5.5 cm for BeiDou, using real-time corrections for clocks and satellite orbits [48].
[2] Operational world or regional GNSS: GPS (United States of America), GLONASS (Russia), BeiDou (China), Galileo (European Union), NavIC (India), QZSS (Japan).

- Uncertainty quantification in measurement systems: Measurement uncertainty is an inherent challenge in any engineering system, especially one that integrates many advanced technologies. For smart lenses, this uncertainty could affect a variety of measurements, such as light intensity, optical alignment, sensor readings, and even user interaction. Quantifying this uncertainty helps engineers design systems that account for potential errors and variability in performance. By applying uncertainty quantification techniques, such as statistical analysis and error modeling, developers can ensure that their smart lenses provide reliable, consistent performance even in the face of manufacturing tolerances, environmental variations, or wear and tear over time. This is essential for ensuring the accuracy of measurements and the overall effectiveness of the device. We rely on our experience in calculating uncertainties on complex microwave optics systems [53–56].

- Patent landscape awareness to secure intellectual property and funding: Intellectual property (IP) is a key asset when developing innovative technologies such as smart lenses [57,58]. A deep awareness of the patent landscape allows developers to protect their innovations from competitors and ensure that their work doesn't infringe on existing patents. Additionally, a strong IP portfolio can be essential for securing funding from investors or institutions, as it adds credibility to the project and demonstrates a clear competitive advantage. The patent landscape for smart lenses likely includes patents for the lens design itself, the technologies integrated into the lenses (e.g., microelectronics, sensors, wireless communication), and methods for manufacturing these lenses. Staying ahead of trends in patent filings and securing broad protection for novel innovations can be a strategic advantage.

These domains must converge to develop viable, safe, and functional prototypes of energy-autonomous smart lenses. The development of energy-autonomous smart lenses requires the convergence of all these areas of expertise. Successful integration of thin-film coatings, biocompatible materials, optoelectronic systems, and robust signal diagnostics must occur within a framework of intellectual property management and careful measurement of uncertainties. For example, a lens that uses optical coatings to enhance visual quality must also have embedded sensors or electronics that are seamlessly powered and communicate with external devices without interference or signal degradation. The integration of these technologies is not only about functionality but also about user safety and usability. The ultimate goal is to create a viable prototype that is not only functional but safe to use, efficient in power consumption (hence autonomous), and manufacturable at scale. Creating smart lenses is a multidisciplinary endeavor where expertise in optics, materials science, electronics, patent law, and system engineering must be harmonized to ensure the development of safe, high-performing, and commercially viable products.

## 5. Operating frequency range

This section presents the technical characteristics of a compact stacked double loop antenna designed for directional electromagnetic transmission and reception. The loop antenna or magnetic loop antenna is sensitive to the magnetic field (hence its name magnetic loop). Its operating principle results from a direct application of the Lenz-Faraday law, the induced voltage being proportional to the flux of the magnetic field, the loop antenna belongs to the category of fluxmeters. We can refer to theoretical consideration although antenna are keys components in telecommunication, control and data acquisition [59–61].

The structure comprises two vertically aligned circular loops, which function collectively as magnetic dipoles. Such a configuration enhances directional radiation, making it suitable for applications requiring focused signal propagation or sensing. The antenna has a loop diameter of 9.5 mm. It is chosen to stay compatible with typical dimensions of lenses and without masking the vision. We will see that it corresponds to a resonant frequency near 1.01 GHz, based on established principles of small loop antenna theory. The effective operating frequency range spans from 900 MHz to 1.1 GHz, covering a bandwidth appropriate for many modern communication and detection systems. The antenna's radiation pattern is optimized along the vertical axis (broadside to the loop plane), where emission and reception are strongest at angles $\theta = 0°$ and $\theta = 180°$. This directionality improves signal gain and precision in targeted scenarios. This section outlines the relevant physical parameters, theoretical calculations, and practical considerations including the use of the speed of light in estimating frequency and wavelength. This analysis contributes to a broader understanding of antenna performance, design constraints, and application-specific deployment within the specified GHz range.

*5.1. Antenna Structure, Dimensions, Range and Resonant Frequency*

A stacked double loop antenna consists of two circular loops aligned vertically, functioning as magnetic dipoles. This structure is often used for directional emission and reception, behaving like two magnetic dipoles in phase. These antennas are used for directional radiation and reception.

For a loop with a 9.5 mm diameter (D = 0.0095 m), the radius is r = D / 2 = 4.75 mm, leading to a loop circumference of approximately $C = 2\pi r = 0.0298$ m. In small loop antenna theory, resonance typically occurs when the circumference is about one-tenth of the wavelength. Thus, the estimated resonant wavelength is 0.298 m.

For small loop antennas, the resonant condition typically occurs when the loop circumference is approximately 1/10 of the wavelength ($\lambda$). We assume that the loop diameter is D = 9.5 mm. The key criterion for a small loop antenna is: loop circumference C is negligible compared to $\lambda$ with c / $\lambda$ < 1/10.

Using those formula:

- $\lambda \approx 10 \times C = 0.298$ m

- $f = c / \lambda \approx 3 \times 10^8 / 0.298 \approx 1.01$ GHz,

and using the speed of light (c = $3 \times 10^8$ m/s), the resonant frequency is approximately determined as 1.01 GHz. Given a centered resonant frequency of ~1.01 GHz, the typical usable bandwidth is approximately 900 MHz to 1.1 GHz. Since antennas are reciprocal, they operate similarly for both transmission and reception, that's why the expected operating frequency range for this antenna is in [900 MHz, 1.1 GHz]. The exact usable bandwidth may vary depending on the antenna's Q-factor and matching components.

Since antennas are reciprocal, the operating frequency range is the same for both transmission and reception. Given a centered resonant frequency of ~1.01 GHz, the typical usable bandwidth is approximately 900 MHz to 1.1 GHz.

*5.2. Directional Properties, Preferred Angles and Q-factor*

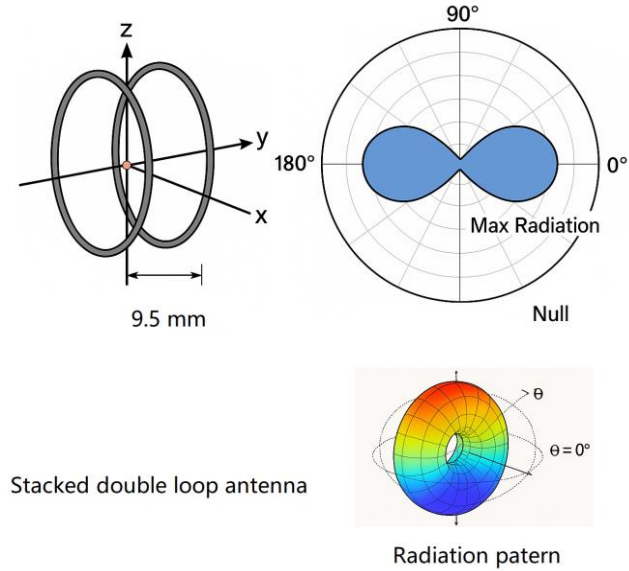

**Figure 3.** Small double loop antenna of 9.5 mm diameter, real distances between antenna are not respected to enable a better representation. Radiation of a small loop antenna is donut-shaped like a magnetic dipole. Maximum radiation is in the plane of the loop. Color for illustration purpose. Null – or Zero radiation, is along the axis perpendicular to the loop plane. For small loops, radiation resistance $R_r = 20\pi^2 \times (A/\lambda^2)^2$ where $A = \pi r^2$ is the area of the loop and $\lambda = c/f$.

Figure 3 gives the small double loop antenna of 9.5 mm diameter main characteristics. The radiation pattern of a single loop is Donut-shaped with maximum emission perpendicular to the plane of the loop. However, in a stacked configuration, the pattern becomes more directional, with maximum radiation occurring along the vertical axis (broadside), i.e., at $\theta = 0°$ and $\theta = 180°$ relative to the loop plane. So preferred emission/reception directions are along the axis of the loop stack: $\theta = 0°$ and $\theta = 180°$. This improves directivity and gain in those directions, making the antenna more effective for targeted communication or sensing. When two such loops are stacked, the radiation is reinforced along the axis of the loop stack. Single loop antenna radiates most strongly in the plane perpendicular to its axis. The Q-factor (quality factor) of a small loop antenna is typically estimated as $Q \approx f_0 / \Delta f$. Given $f_0 = 1.01$ GHz and $\Delta f = 200$ MHz (-3 dB bandwidth), the estimated Q-factor is approximately $Q \approx 1001 / 200 = 5.05$. This is very low, which is favorable for wideband operation and suggests the antenna is efficient for communication.

We can also roughly estimate the theoretical Q-factor of a small loop antenna using the radiation resistance $R_r$ and loss resistance $R_l$: $Q = 2\pi f L / R$. However, this requires an estimation of the inductance L of the loop, and knowing or approximating the resistance R. For a single-turn circular loop of radius r, made from a thin conductor, inductance of the loop is $L = \mu_0 r(\ln(8r/a) - 2)$ where r=4.75 mm, wire radius could be 10 μm, and $\mu_0 = 4\pi \times 10^{-7}$ H/m. Using this, we can say the inductance is in the nH range, and combined with expected loss resistance of several ohms, a Q-factor of 5 to 10 is typical and reasonable for such compact antennas.

*5.3. Discussion*

Uncertainty has to be systematically estimated as we already discuss in part 4 of this manuscript [53–56] and it is important to underline that with such system typically operating from 900 MHz to 1.1 GHz. The upper limit of 1.1 GHz is the physical limit for small loop behavior for 9.5 mm loops. This frequency range is compatible with control operated by compact and portable circuits. In this section we summarize the main parameters for the operating frequency range and direction of antennas. They are given in Table 2. These settings are perfectly compatible with transmission to a portable device such as a smartwatch or smartphone

**Table 2.** Settings for the operating frequency range and direction of antennas.

| Parameter | Determination | Value – Range |
|---|---|---|
| Operating Frequency[1] | Tx/Rx | 900 MHz – 1.1 GHz |
| Resonant Frequency Estimation | $f = c / \lambda$ | 1.01 GHz |
| Preferred Radiation Angles | | 0° – 180° |

[1] Wavelength estimation: $\lambda \approx 10C$ with small loop approximation.

# 6. Conclusion

This paper presents a comprehensive overview of the technological, biomedical, and electromagnetic considerations involved in the development of smart contact lenses with autonomous power supply. The integration of energy-harvesting mechanisms, compact antennas, and biocompatible components within the highly constrained form factor of a contact lens presents significant challenges—but also remarkable opportunities for next-generation wearable devices. A central focus of this study is the exploration of energy sources intrinsic to the human eye. Tear-based cells leverage the natural salinity to generate electricity, offering a continuous and safe alternative to conventional batteries. Meanwhile, biomechanical energy from blinking demonstrates the feasibility of converting frequent eyelid motion into usable power. These two strategies are complementary and scalable, making them highly promising for long-term, real-world applications.

To enable wireless data communication and system control, we proposed a stacked double-loop antenna operating within the 900 MHz to 1.1 GHz range. With a 9.5 mm diameter, this antenna fits the geometry of a soft contact lens and ensures efficient directional transmission toward portable devices such as smartphones and smartwatches. This specific bandwidth was chosen due to its compatibility with existing mobile and wireless standards, allowing seamless communication with devices like smartphones and smartwatches. The proposed antenna design maximizes radiation efficiency along preferred vertical axes ($\theta = 0°$ and $\theta = 180°$), and exhibits a low Q-factor (~5), supporting wideband data transmission while fitting comfortably within the spatial limits of a soft contact lens. These characteristics are essential for enabling bidirectional communication. Beyond energy and signal transmission, the design must also ensure biocompatibility, transparency, oxygen permeability, and user comfort. Material selection and the need to comply with specific absorption rate (SAR) limits were thoroughly discussed. The review of international SAR regulations (e.g., IEC/IEEE guidelines) and wireless communication underscore the need to apply precautionary principles and rigorous validation during product development. We also addressed engineering requirements such as signal integrity, measurement uncertainty, and manufacturing scalability via microfabrication and 3D printing techniques. Our prior work in metrology and optics provided a foundation for evaluating these performance metrics, including modeling noise, intermodulation effects, and thermal stability in microelectronic subsystems. The pathway toward viable, self-powered smart lenses is becoming clearer. The integration of bioenergy sources, compact RF systems, and adaptive optical materials suggests exciting potential for applications in health monitoring, vision enhancement, and real-time user interaction. Continued interdisciplinary collaboration is a key to transforming these prototypes into safe, functional, and market-ready devices.

**Dedicated** to the memory of Pr. Robert Gabillard (1926 – 2012), who would soon have been 100 years old, an antenna expert, pioneer in the use of antennas in new disciplinary fields, who has made his mark in the fields of underground telecommunications and automated transport. He contributed to geophysical detection of gas and oil, the Lille metro (VAL), the Channel Tunnel between France and England, and Line 14 of the Parisian automated metro, combining academic research, industrial innovation, and regional planning.

**Supplementary Materials:** Not applicable.

**Author Contributions:** Conceptualization, P.S. and M.C.; methodology, P.S.; figures, B.G. and P.S.; validation, P.S.; formal analysis, P.S.; investigation, P.S.; functional principles, P.S. and E.P.; 3D-printing lenses investigation, A.M.B.; resources, P.S. and M.V.P.; writing—original draft preparation, P.S.; writing—review and editing, P.S.; supervision, P.S.; project administration, P.S.; funding acquisition, P.S. All authors have read and agreed to the published version of the manuscript.

**Funding:** This research received no external funding.

**Data Availability Statement:** Data is available upon reasonable request.

**Acknowledgments:** This work was made possible in particular by the National Centre for Scientific Research (CNRS), a government-funded research organization under administrative authority of France's Ministry of research, which allows the creation of multidisciplinary teams to explore new research topics in the field of Micro Nano Sciences and Systems.

**Conflicts of Interest:** The authors declare no conflicts of interest.

## Abbreviations

The following abbreviations are used in this manuscript:

| | |
|---|---|
| CNRS | Centre National de la Recherche Scientifique |
| UMLP | Université Marie et Louis Pasteur |
| FEMTO-ST | Institut Franche-Comté Mécanique Thermique Optique Sciences et Technologies |
| ITMO | University of Information Technologies, Mechanics and Optics |
| CLP | Contact Lens Pointer |
| SCL | Smart Contact Lens |
| e.g. | For example, or *exempli gratia* in Latin |
| CMOS | Complementary Metal-Oxide-Semiconductor |
| PDMS | Polydimethylsiloxane |
| ASIC | Application-specific integrated circuit |
| SAR | Specific Absorption Rate |
| ICNIRP | International Commission on Non-Ionizing Radiation Protection |
| PB | Prussian Blue, or *Preußischblau* in German |
| mAh | milliampere-hours |
| DC | Direct Current |
| IP | Intellectual Property |
| Wi-Fi | Wireless Fidelity |
| Mbps | megabits per second |

| Gbps | gigabits per second |
| 3G | third generation of wireless mobile |
| 4G | fourth generation wireless mobile |
| 5G | fifth generation wireless mobile |
| GNSS | Global Navigation Satellite System |
| GPS | Global Positioning System |
| GLONASS | GLObalnaya NAvigatsionnaya Sputnikovaya Sistema |
| BeiDOU | BeiDou is Chinese for the Big Dipper |
| NavIC | Navigation with Indian Constellation |
| QZSS | Quasi-Zenith Satellite System |